\documentclass{elsart}
\journal{Solid State Communications}

\usepackage{graphicx}
\usepackage{dcolumn}
\usepackage{rotating}
\usepackage{amssymb}
\usepackage{amsmath}
\usepackage{float}
\usepackage{color}

\begin{document}

\begin{frontmatter}



\title{Seebeck coefficients of half-metallic ferromagnets}
                         
\author[Mainz]{Benjamin~Balke,}
\ead{balke@uni-mainz.de}
\author[Mainz]{Siham Ouardi,}
\author[Mainz]{Tanja Graf,}
\author[Mainz]{Joachim Barth,}
\author[Mainz]{Christian G. F. Blum,}
\author[Mainz]{Gerhard~H.~Fecher,}
\author[EMPA]{Andrey Shkabko,}
\author[EMPA]{Anke Weidenkaff,}
\author[Mainz]{Claudia~Felser}

\address[Mainz]{Institut f\"ur Anorganische Chemie und Analytische Chemie,\\
Johannes Gutenberg-Universit\"at Mainz,
D-55099 Mainz, Germany}
\address[EMPA]{EMPA, Swiss Federal Laboratories for Materials Testing and Research, \\
Solid State Chemistry and Catalysis,
CH-8600 Duebendorf, Switzerland}

\begin{abstract}
In this report the Co$_2$ based Heusler compounds are discussed as potential materials for spin voltage generation. The compounds were synthesized by arcmelting and consequent annealing. Band structure calculations were performed and revealed the compounds to be half-metallic ferromagnets. Magnetometry was performed on the samples and the Curie temperatures and the magnetic moments were determined. The Seebeck coefficients were measured from low to ambient temperatures for all compounds. For selected compounds high temperature measurements up to 900~K were performed. 
\end{abstract}

\end{frontmatter}

\section{Introduction}

In the recent years Heusler alloys have attracted a lot of interest
as suitable materials for spintronic applications~\cite{FFB07}.
A huge amount of studies investigating the half-metallic properties
theoretically and experimentally and enhancing the performance
of the compounds and devices for different type of applications were
done. In 2005, Hashemifar {\it et al}~\cite{HKS05} showed in a
density functional theory study that it is possible to preserve
the half-metallicity at the Heusler compound Co$_2$MnSi(001) surface
and very recently, Shan {\it et al}~\cite{SSW09} demonstrated experimentally
the half-metallicity of Co$_2$FeAl$_{0.5}$Si$_{0.5}$ at room temperature.
The ability of growing well ordered half-metallic Heusler thin films on
semiconductors (e.g. Ge (111)) makes Heusler compounds suitable
for spin injection as shown very recently by Hamaya {\it et al}~\cite{HIN09}.
The recent observation of the spin Seebeck effect
allows to pass a pure spin current over a long distance \cite{UTH08}
and is directly applicable to the production of
spin-voltage generators which are crucial for driving spintronic devices~\cite{WAB01,ZFS04,CFV07}.
For an effective generation of a spin current the spin Seebeck coefficient and the spin voltage namely the difference in the chemical potential of the spin up $\mu_{\uparrow}$ and spin down $\mu_{\downarrow}$ should be large~\cite{Ong08}. In this report Co$_2$ based Heusler compounds  are investigated on their potential as spin voltage generators. The high application potential of the compounds is demonstrated. For a further optimization the electronic and magnetic properties of the compounds can be designed easily.

\section{Experimental Details}

The samples were produced by arc melting of stoichiometric amounts of the
elements and afterwards annealed in evacuated quartz tubes for 21~days.
For more details about the sample preparation see Ref.~\cite{BFK06}.
The magnetic properties of the samples were investigated by a super
conducting quantum interference device (SQUID, Quantum Design
MPMS-XL-5).
The measurements of the Seebeck coefficient were carried out with
a Physical Property Measurement System (Model 6000 Quantum Design)
on bars of about $(2\times2\times8)$~mm$^3$ which were cut from the pellets and polished
before the measurement. In the temperature range above 350~K the Seebeck coefficient was measured by a steady-state method using the RZ2001i measurement system from Ozawa science, Japan.
For more details about the preparation, structural and magnetic properties see References ~\cite{WFK05,OGB09,GFB09,BFB09}.


\section{Results and Discussion}
\subsection{Calculational Results}
\begin{figure}
  \centering
  \includegraphics[width=12cm]{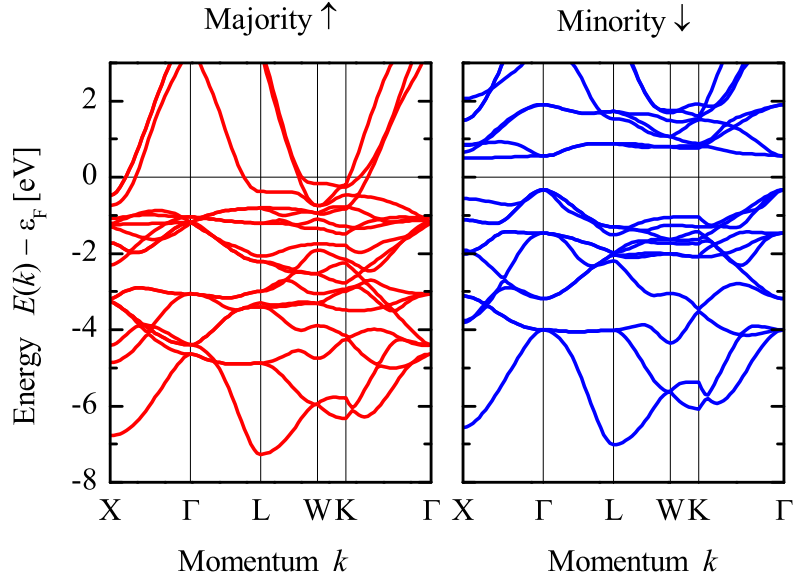}
  \caption{Band structure of Co$_2$MnSi.\newline
         The band structure of Co$_2$MnSi is displayed for the majority and the minority channel.}
  \label{Fig1_BS-Co2MnSi.EPS}
\end{figure}

As a starting point, the electronic structure
of the compounds was calculated using 
the FLAPW method. The detailed description of the calculations for Co$_2$YZ~\cite{KFF07c} can be found in~\cite{GFB09,KFF07c}.
The investigated compounds are displayed in Table~\ref{table_data}. 
In spin polarized calculations the compounds turn out to be half-metallic 
ferromagnets as was previously shown by various
authors~\cite{MAH06,KFF07,IAK82,GDP02,LLB05}. As an example the majority and minority band structure of Co$_2$MnSi is displayed in Figure~\ref{Fig1_BS-Co2MnSi.EPS}.

\begin{figure}[H]
  \centering
  \includegraphics[width=12cm]{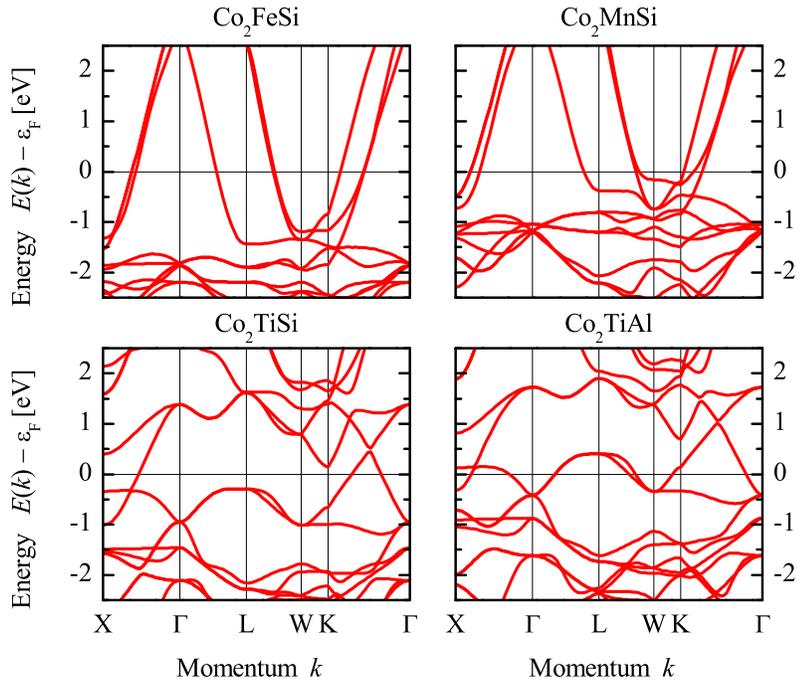}
  \caption{Majority band structure calculations for selected compounds.\newline
         Displayed are the majority bands for
         Co$_2$TiAl, Co$_2$TiSi, Co$_2$MnSi, and Co$_2$FeSi. For a better comparison
         only the energy range $\pm~2.5~eV$ around $\epsilon_F$ is shown.}
  \label{Fig2_Spag-all}
\end{figure}

For selected Heusler compounds the majority band structures are shown in Figure~\ref{Fig2_Spag-all}. For a better comparison
only the energy range $\pm~2.5~eV$ around the Fermi energy $\epsilon_F$ is shown. One can clearly see the difference between
Co$_2$MnSi and Co$_2$FeSi (upper panels) and Co$_2$TiSi and Co$_2$TiAl (lower panels). In the case of 
Co$_2$MnSi and Co$_2$FeSi the bands which cross $\epsilon_F$ are strongly dispersive 
while the bands are almost flat around  $\epsilon_F$ for Co$_2$TiSi and Co$_2$TiAl.
The minority bands are not shown here because they look almost the same for almost all compounds
with a gap around $\epsilon_F$. Furthermore, the minority electrons don't contribute to the transport properties for not too high
temperatures. This minority gap is characteristic for half-metallic materials. The occurrence of the gap in one spin direction 
is a fundamental prerequisite for materials used as spin voltage generators.

\subsection{Experimental Results}

Table~\ref{table_data} summarizes the measured data on the various compounds that will be
discussed below in detail. The compounds are grouped by their number of valence electrons.

\begin{table}[H]
\caption{Magnetic moments, Curie temperatures, and Seebeck coefficients of the investigated Co$_2$YZ compounds.}
\smallskip
\centering\begin{tabular}{c c c c c}
            \hline
            Compound    & $T_c$ [K] & $M_{sat}(5~K)$ [$\mu_B$]   & $S$(300~K)[$\mu$VK$^{-1}$]   \\
            \hline
            Co$_2$TiAl  &  128      & 0.75                        &   -55                      \\
            \hline
            Co$_2$TiSi  &  380      & 1.96                        &   -27                       \\
            Co$_2$TiGe  &  380      & 1.94                        &   -22                     \\
            Co$_2$TiSn  &  355      & 1.97                        &   -34                      \\
            \hline
            Co$_2$MnAl  &  693      & 3.96                        &   -4                        \\
            \hline 
            Co$_2$MnSi  &  985      & 4.97                        &   -7                        \\
            Co$_2$MnGe  &  905      & 4.98                        &   -15                     \\
            Co$_2$MnSn  &  829      & 5.03                        &   -33                      \\
            \hline
            Co$_2$FeSi  &  1100     & 5.97                        &   -12                       \\
            \hline     
            
\end{tabular}
\label{table_data}
\end{table}

The Slater-Pauling curve~\cite{Sla36,Pau38} is a simple way to study
for ferromagnetic alloys the interrelation between the valence electron
concentration and the magnetic moments. It is
well known that Heusler compounds based on Co$_2$ follow the
Slater-Pauling rule for predicting their total spin magnetic moment
\cite{Kue00,GDP02,FKW06} that scales linearly with the number of
valence electrons. 
In the case of four atoms per primitive cell, as in Heusler compounds, 
one has to subtract 24 (for more details see Ref.~\cite{KFF07} from the accumulated number of
valence electrons in the primitive cell $N_V$ ($s, d$ electrons for the
transition metals and $s, p$ electrons for the main group element) to
find the magnetic moment per cell ($m$):

\begin{equation}
       m = N_V - 24,
\label{eq1}
\end{equation}

with $N_V$ denoting the accumulated number of valence electrons in the
primitive cell. In the case of Heusler compounds, the
number 24 arises from the number of completely occupied minority bands
that has to be 12 in the half-metallic state. In particular these are
one $s$ ($a_{1g}$), three $p$ ($t_{1u}$), and eight $d$ bands
\cite{Kue00,Kue84}. The later consist of two triply degenerate bands
with $t_{2g}$ symmetry and one with $e_g$ symmetry (note that the given
assignments of the irreducible representations are only valid at the
$\Gamma$-point and neglecting the spin of the electrons.).

The measured values are displayed in Table~\ref{table_data} 
and almost all the compounds follow the Slater-Pauling behaviour very well
which is a precondition for half-metallic ferromagnetism. The magenetic moment
of Co$_2$TiAl is lower than the expected of 1~$\mu_B$, this is explaind by
partial disorder (for details see Ref.~\cite{GFB09}).

\begin{figure}
  \centering
  \includegraphics[width=12cm]{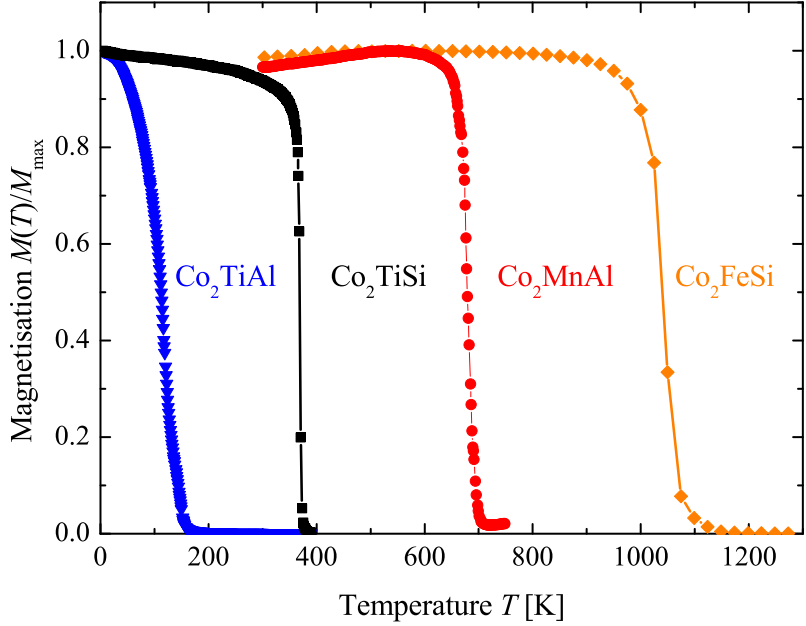}
  \caption{Temperature dependent magnetisation of Co$_2$TiAl, Co$_2$TiSi, Co$_2$MnAl, and Co$_2$FeSi.
  				The data are normalized for better comparison. The magnetisation was measured in a low induction
  				field to make the phase transition clearly visible. This causes that the shape of the curves differ
  				from the molecular filed-like behaviour.}
  \label{Fig3_Tc-all}
\end{figure}

Figure~\ref{Fig3_Tc-all} shows the measurements of the temperature dependent 
magnetisation of Co$_2$TiAl, Co$_2$TiSi, Co$_2$MnAl, and Co$_2$FeSi. From this kind of 
measurements the Curie temperatures were distinguished, the values are summarized in Table~\ref{table_data}.
With the increase of the number of valence electrons and therefore as well the
increase of the magnetic moment the Curie temperature increases. 
One can nicely see the increase of the Curie temperature for the samples
with higher numbers of valence electrons. These behaviour is well known for Heusler
compounds and was reported already some years ago~\cite{FKW06,KFF07b}.
This measurements show the easy tunability of the magnetic properties of the
Heusler compounds. By controlling the number of valence electrons one can design
the properties of the materials.

\begin{figure}
  \centering
  \includegraphics[width=12cm]{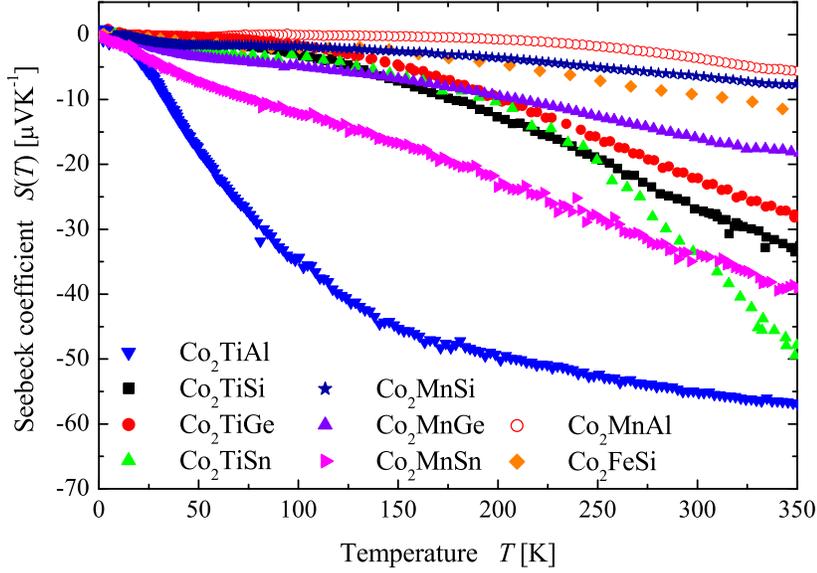}
  \caption{Measured Seebeck coefficients (\textit{S}) for selected Heusler compounds from 2~K to 350~K.}
  \label{Fig4_Seebeck-LT}
\end{figure}

In Figure~\ref{Fig4_Seebeck-LT} the measured Seebeck coefficients (\textit{S}) are displayed
for temperatures from 2~K to 350~K. 
The absoule values increase with increasing temperature. The values at 300~K of all the measurments 
are summarized in Table~\ref{table_data}. For an increase of the valence electron count 
the absolute value of the Seebeck coefficient is decreased. This is explained by the increase 
of the electron concentration in the bands. By increasing the valence electron count additional 
electrons are added to the d-band at the Fermi energy~\cite{OBG09}. This leads to an increase 
of the carrier concentration. The increase of the carrier concentration leads to a decrease 
of the Seebeck coefficient~\cite{SnT08}. The interrelationship between carrier concentration 
and Seebeck coefficient can be seen from relatively
simple models of electron transport. For simple metals or degenerate
semiconductors with parabolic bands and energy-independent scattering the Seebeck coefficient \textit{S} is given by

\begin{equation}
       S = \frac{8 \pi^2 k^{2}_{B}}{3eh^2} m^* T \left(\frac{\pi}{3n}\right)^{3/2},
\label{eq2}
\end{equation}

where \textit{n} is the carrier concentration and \textit{m*} is the effective mass of
the carrier. It can be clearly seen that \textit{S} dependens on the carrier concentration 
and on the effective mass \textit{m*}. The latter depend on the shape of the bands. Although, 
the Seebeck coefficient is decreasing with increasing carrier concentration it is remarkable 
that Sn containing compounds yield large absolute values for the Seebeck coefficient. 
This effect is related to changes in the band structure and consequential changes in the 
effective mass \textit{m*}. The Seebeck coefficient of the investigated compounds is 
negative over the entire temperature range. The found absolute values are quite large 
compared to elemental metals ($S_{Co}$=-30.8, $S_{Mn}$=-9.8, $S_{Ti}$=+9.1 $S_{Al}$=-1.66, $S_{Sn}$=-1, 
values at $T$=300~K and in $\mu$VK$^{-1}$ ). Especially the Sn containing compounds show high absolute 
Seebeck coefficients.

\begin{figure}
  \centering
  \includegraphics[width=12cm]{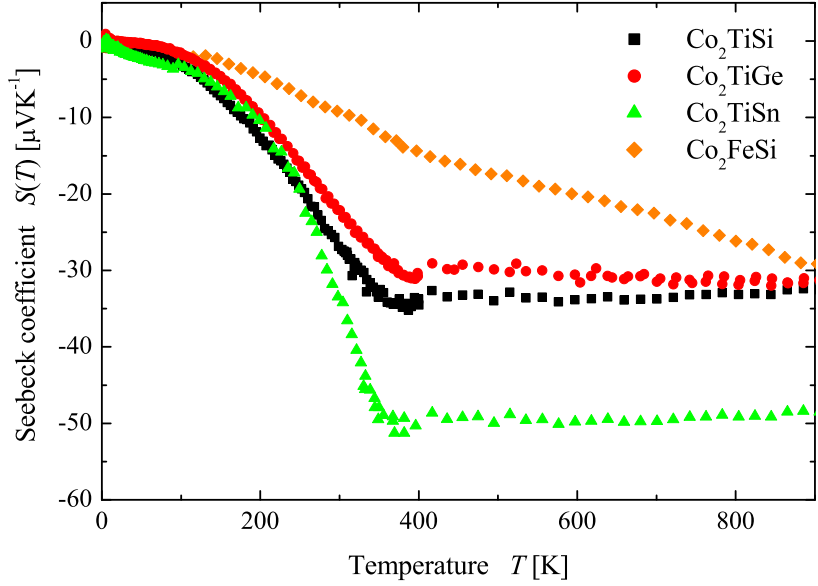}
  \caption{Measured Seebeck coefficients (\textit{S}) for seleted Heusler compounds. 
           Shown are the data for the complete temperature range from 2~K to 900~K for Co$_2$FeSi, Co$_2$TiSi, Co$_2$TiGe, and Co$_2$TiSn.}
  \label{Fig5_Seebeck-HT}
\end{figure}

For selected compounds high temperature measurements of the Seebeck coefficient have been carried out. 
The results are shown in Figure~\ref{Fig5_Seebeck-HT}. At the ferromagnetic to 
paramagnetic transition temperature a significant change in the slope is observed around T$_c$.  
For the compounds Co$_2$TiSi, Co$_2$TiGe, and Co$_2$TiSn a nearly constant behaviour of the Seebeck 
coefficient above T$_c$ is observed~\cite{BFB09}. Although not shown here, measurements of the 
Seebeck coefficient for Co$_2$FeSi have been carried out up to 1300~K. A pronounced peak of 
the Seebeck coefficient has been observed around T$_c$. 

\section{Summary}
Co$_2$ based Heusler compounds have been investigated on their potential use as spin voltage generators. The compounds have been synthesized by arc melting and consequent annealing. Magnetometry was performed and the Curie temperatures and the magnetic moments were determined. The Seebeck coefficient was investigated in the low temperature range from 2~K to 350~K and for selected compounds for higher temperatures. The observed Seebeck coefficients were all negative in the whole temperature range. The absolute values are quite large compared to simple metals. Especially the Sn containing compounds show high Seebeck coefficients with high Curie temperatures. This makes them attractive candidates for materials used in spin voltage generators. 

\section{Acknowledgments}

The authors thank Stiftung Rheinland-Pfalz f{\"u}r
Innovation (Project 863) and the Deutsche Forschungsgemeinschaft DFG (research unit FOR~559) for the financial support of the project.

\bibliographystyle{elsart-num}
\bibliography{Balke_Seebeck_SSC}

\end{document}